\documentclass[final]{cvpr}

\usepackage{times}
\usepackage{epsfig}
\usepackage{graphicx}
\usepackage{amsmath}
\usepackage{amssymb}
\usepackage{tabularx}
\usepackage{booktabs}

\usepackage[pagebackref=true,breaklinks=true,colorlinks,bookmarks=false]{hyperref}

\def\cvprPaperID{6278} 
\def\confYear{CVPR 2021}
\begin{document}

\title{Wearing the Same Outfit in Different Ways -- A Controllable Virtual Try-on Method}

\author{Kedan Li\\
{\tt\small kedan@revery.ai}
\and
Jeffrey Zhang\\
{\tt\small jeff@revery.ai}
\and
Becca Chang\\
{\tt\small shaoyuc3@illinois.edu}
\and
David Forsyth\\
{\tt\small daf@illinois.edu}
}
\maketitle

\begin{abstract}
  An outfit visualization method generates an image of a person wearing real garments from images of those garments. Current methods can produce images that look realistic and preserve garment identity, captured in details such as collar, cuffs, texture, hem, and sleeve length. However, no current method can both control how the garment is worn -- including tuck or untuck, opened or closed, high or low on the waist, etc.. -- and generate realistic images that accurately preserve the properties of the original garment.

  We describe an outfit visualization method that controls drape while preserving garment identity. Our system allows instance independent editing of garment drape, which means a user can construct an edit (e.g. tucking a shirt in a specific way) that can be applied to all shirts in a garment collection. 
  Garment detail is preserved by relying on a warping procedure to place the garment on the body and a generator then supplies fine shading detail. To achieve instance independent control, we use control points with garment category-level semantics to guide the warp.

  The method produces state-of-the-art quality images, while allowing creative ways to style garments, including allowing tops to be tucked or untucked; jackets to be worn open or closed; skirts to be worn higher or lower on the waist; and so on. The method allows interactive control to correct errors in individual renderings too. Because the edits are instance independent, they can be applied to large pools of garments automatically and can be conditioned on garment metadata (e.g. all cropped jackets are worn closed or all bomber jackets are worn closed).
  
\end{abstract}



\begin{figure}
\begin{center}
\vspace{-0.5cm}
	\includegraphics[width=0.98\linewidth]{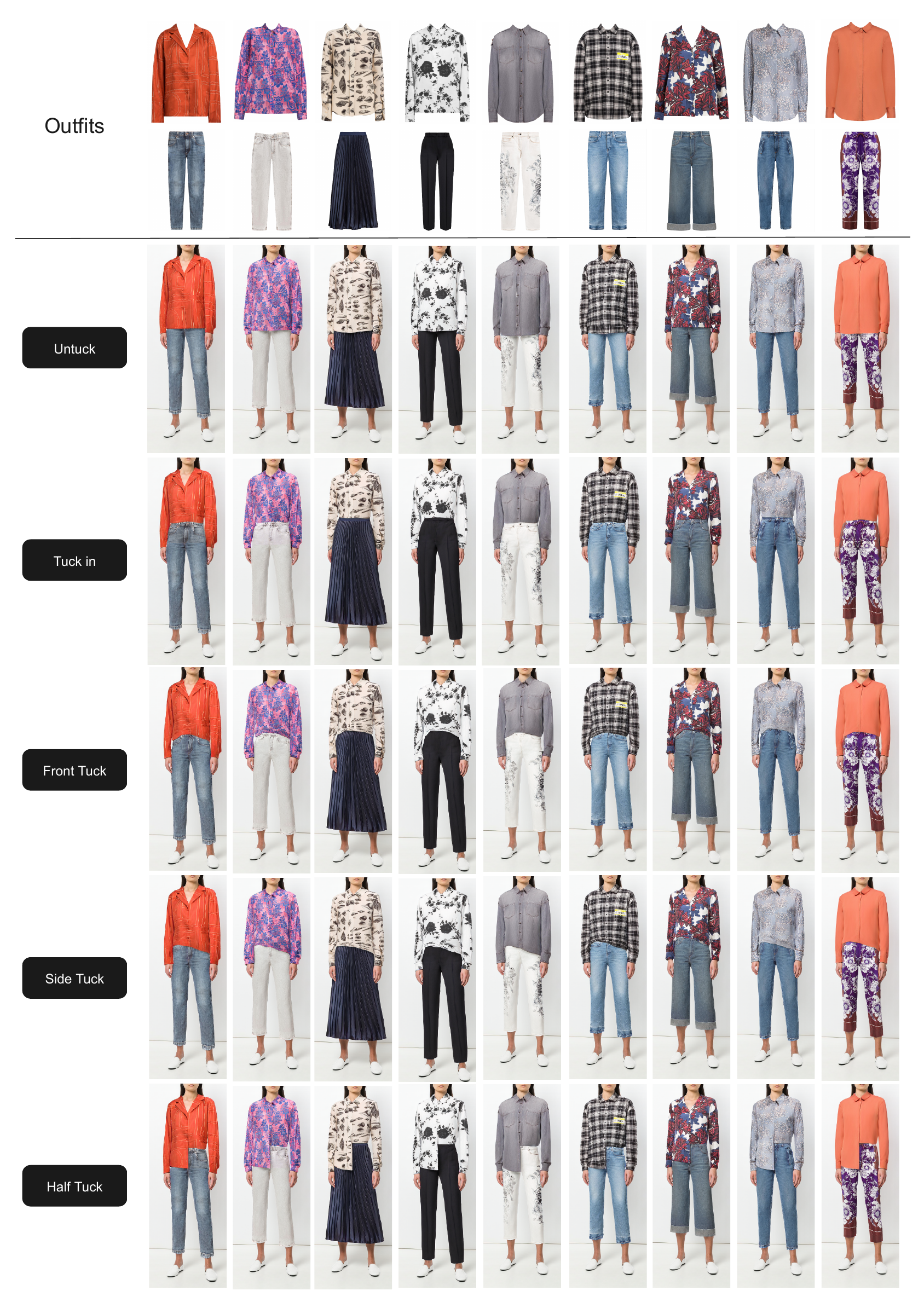}
        \caption{
        Our method produces high quality images of people wearing a provided outfit, while allowing the garments to be worn in different ways. The figure shows our method draping the same shirt untucked or tucked in many different styles. Each {\em column} shows images of the same garment worn in different ways and each {\em row} shows different outfits worn in the same style/drape. As shown, the same kind of drape renders well on all shirts and the identity of the garments is unaltered when worn in different ways.
        }    
    \label{fig:different_tuck}
    \end{center}
\vspace{-0.9cm}  
\end{figure}

\section{Introduction}

Current virtual try-on (VTON) methods can synthesize compelling images of people wearing a prescribed set of garments, but cannot vary the way in which the garment is worn (the {\em drape} of the garment). However, people wear the same garment in different ways, (e.g. a shirt could be tucked or untucked). Each {\em row} of Figure~\ref{fig:different_tuck} shows images of different garments worn with the same drape and synthesized by our system. The alternatives could be discrete (a jacket could be worn open or closed) or continuous (a skirt worn at different points on the waist or hips) and are often mixed (tucked shirts largely look the same, but there are many ways to wear a shirt untucked). A system that allows a user to see the same garment in different drapes must meet three constraints. First, the system must {\em preserve} garment identity, so that however the garment is shown, it is always the same garment.  Each {\em column} of Figure~\ref{fig:different_tuck} shows images synthesized by our system of the same garment worn in different ways. Second, the system must allow {\em  instance independent control}, so that (for example) the same edit can be applied to
all t-shirts or all outerwear. Third, the generated images must be {\em realistic} by accurately generating the garments while showing natural interactions between garments and body.  We describe a method to produce instance independent edits that preserve garment identity and result in high-quality images.


There is now a substantial amount of literature on VTON methods that produce highly realistic
images~\cite{Dong2018SoftGatedWF,Lassner:GeneratingPeople:2017,hsiao2019fashionplus,zhu2017be,Han_2019_ICCV,wang2018toward,Rocco17,tprvton,Issenhuth2020DoNM,  ge2021parser, cycle_consistency_tryon, choi2021vitonhd, Chopra_2021_ICCV, Style_Based_Global_Appearance,  Kedan_Li_2021_CVPR, Neuberger_2020_CVPR}.    
Methods rely on adversarial losses to ensure realistic images.  The core application is selling garments to users, which means preserving garment identity is essential (otherwise users may return garments to vendors). The key to preserving garment identity appears to be to warp the garment from a reference image. The resulting draft image can then be lightly modified by an adversarially trained renderer. However, VTON methods do not yet provide a reliable way to control how a garment is worn on a body.  The problem is difficult because methods must distinguish between image
edits that change the drape (which are desirable) and those that change the garment (which are not). Current methods that achieve fashion edits do not preserve garment attributes ~\cite{zhu2017be,   Raj2018SwapNetIB, hsiao2019fashionplus, Dong_2020_CVPR, men2020controllable, cui2021dressing, Chen_2021_ICCV}.  

Our pipeline is shaped by the three constraints.  To preserve garment detail, our pipeline warps garments to obtain a draft image. To achieve instance independent control, our method specifies warps using control points with category-level garment semantics so that an edit has the same ``meaning'' for each garment of a category.   For example, each of the edits of Figure~\ref{fig:different_tuck} was obtained by editing one
shirt, then pushing the edit to other shirts.  To produce realistic images, our system applies an adversarially trained renderer to the draft image, so producing improved shading, cloth folds, and similar effects.

Each of our control points marks the location of an anchor (e.g., left shoulder, right inner sleeve,
etc..) that is meaningful for the garment category and that
indicates how a garment should drape on a body. It is essential that control points have category-level semantics
so that they can control all instances of the same category. For example, one can apply a slightly tilted tuck to
all shirts by moving the waistline control point of one instance slightly higher, then applying the same
offset to other shirts.  As Figure~\ref{fig:different_tuck} demonstrates, this generalization procedure is successful.
Templates created with this instance independent editing procedure can be triggered by garment metadata, allowing
fashion professionals to create rich style visualization rules which can be applied to entire garment catalogs.
For example, one may want to show all t-shirts as tucked, all the sweaters as untucked, and all the button shirts as half-tucked.

In a typical VTON pipeline, the way garments interact with the body is largely determined by the semantic
layout and the garment warp. We ensure the control mechanism is effective by making the rest of the VTON pipeline
respect the changes made to the control points. We ensure both the layout and the warp respect the control points by
training the semantic layout generator and the warper to be conditioned on the control points. During inference, we
first predict control point locations, then apply any edit template.  The edited control points are then
passed to the semantic layout generator and the warper to obtain the layout and the warp. Finally, the image generator
produces a final image from the layout and warp.  As
shown in Figure~\ref{fig:continuous_split}, the garments in the output image closely follow the control points. 

\begin{figure}
\begin{center}
\vspace{-0.2cm}
	\includegraphics[width=0.98\linewidth]{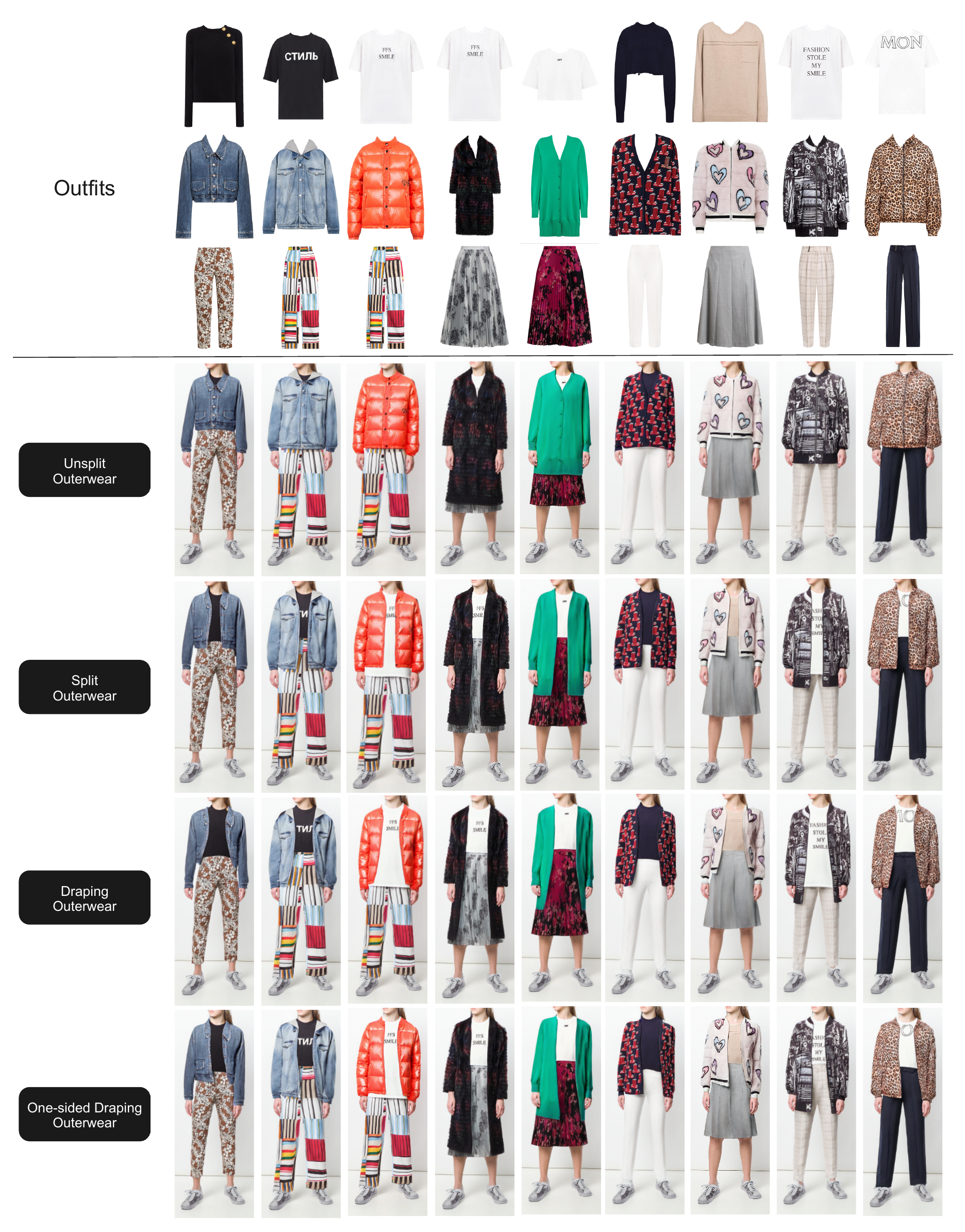}
    \caption{The figure shows a sequence of outfits with outerwear styled differently. We could wear the same jacket as split (unzip or unbuttoned) or non-split (zipped or buttoned) and render it with different drapes. Note that we could produce the same kind of drape for outerwear that are very different (some are knee length while others are cropped). Also, our method is able to preserve very complex patterns in some of the jackets.} 
    \label{fig:different_outerwear}
    \end{center}
\vspace{-0.9cm}  
\end{figure}

\begin{figure}
\begin{center}
	\includegraphics[width=0.98\linewidth]{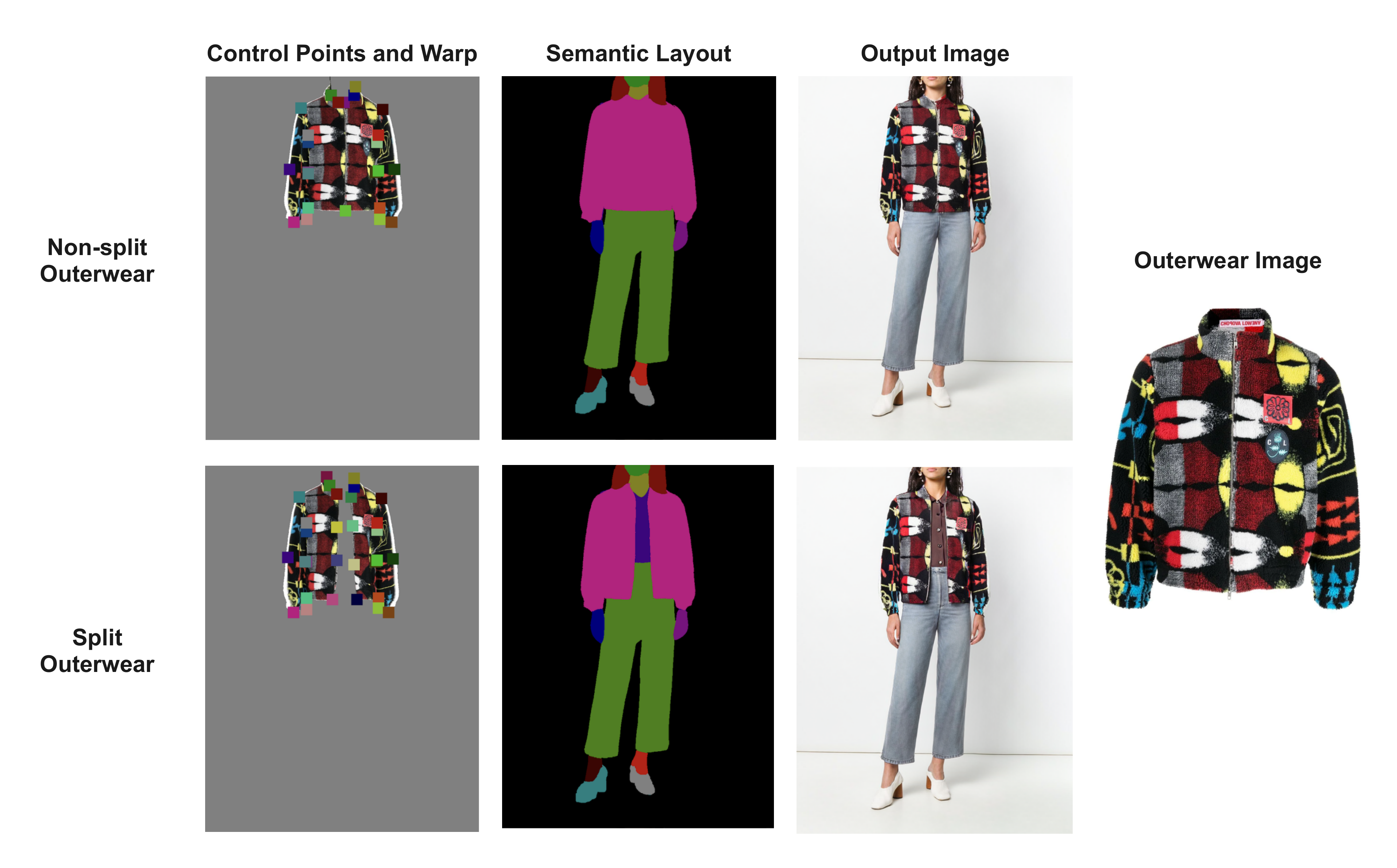}
    \caption{The figure shows examples of the same outerwear worn zipped or unzipped. Note that we use a slightly different set of control points to signal an open jacket. Both the computed warp and the predicted layout exactly follow the control points. The neutral garment is separated into two pieces and warped separately. }    
    \label{fig:outerwear_split}
    \end{center}
\vspace{-0.7cm}  
\end{figure}

Our control mechanism allows a user or a fashion professional
to perform various kinds of edits simply by choosing the set of control points and
moving them around. For example, a jacket can be worn open or closed, by enabling or disabling a few control points as
in Figure~\ref{fig:outerwear_split}. When the jacket is open, it could drape in many possible ways by dragging the
control points on two sides of the opening, as shown in Figure~\ref{fig:continuous_split}. 
In some cases, the control mechanism even allows us to address certain failure cases in prior work due to a lack of coordination between garments, as shown in Figure~\ref{fig:outerwear_skirt}. 

{\bf Contributions:}
We describe the first outfit virtual try-on method that allows reliable control of how a garment is worn on a body without changing the identity of the garment.  We demonstrate  our method can synthesize images of the same garment worn in different and natural ways.  Finally, our control is instance independent: one can make edits on a specific garment and apply the same edit to all the other garments of the same category successfully.

%

\section{Related Work}


\subsection{Image-based Virtual Try-on}

The task of image-based virtual try-on is defined as producing an image of a person wearing a garment with the image of the person and the garment provided. The main challenge lies in producing high quality images while preserving the garment's identity faithfully. Modern image generation networks trained with adversaries can produce high quality images but have difficulty preserving the exact geometrical patterns (such as logos, prints, etc..) on the garment. Thus, most of the works use a differentiable warper to align the garment onto the person to preserve the geometrical patterns~\cite{Dong2018SoftGatedWF,Lassner:GeneratingPeople:2017,hsiao2019fashionplus,zhu2017be,Han_2019_ICCV,wang2018toward,Rocco17,tprvton,Issenhuth2020DoNM, ge2021parser, cycle_consistency_tryon, choi2021vitonhd, Chopra_2021_ICCV, Style_Based_Global_Appearance, Fele_2022_WACV, morelli2022dresscode, lee2022hrviton, Bai2022SingleSV}. The warper is usually guided by semantic layouts which are pixel maps that define the region of the garment and the body parts. Finally, an image generator network trained with adversaries is used to produce high quality try-on images.

In prior work, the warper is guided by the semantic layout and the semantic layout cannot be edited easily. Our method enables editing by introducing control points with garment semantics to the VTON pipeline. The control points are used to guide the warper and the semantic layout, thus allowing a user to modify how the garment is worn by editing the control points.

\subsubsection{Image Warping}

Image warping processes apply a spatial transformation to an image, and thus are able to preserve the 2D patterns on a garment. Unlike the traditional process of computing warps~\cite{Heckbertthesis, Bookstein}, recent methods train a neural network to estimate the warps~\cite{NIPS2015_5854,Ji2017DeepVM,lin2018stgan}. Many virtual try-on works innovates on the method of training differentiable warps: CP-VTON~\cite{wang2018toward} first proposed using a differentiable thin plate spline (TPS) warper and TPRVTON~\cite{tprvton} proposed an improved regularization method; ClothFlow~\cite{Han_2019_ICCV} first introduced a flow-based warping method and is improved by subsequent works~\cite{Style_Based_Global_Appearance, Chopra_2021_ICCV, ge2021parser, Bai2022SingleSV}.

We take inspiration from the warping methods as they are demonstrated to be able to preserve garment attributes faithfully. To achieve controllability, we modified the warping procedure to be guided by control points with garment semantics instead of the semantic layout. Interestingly, recently VTON work~\cite{ge2021parser} removed the semantic layout from the pipeline because the predicted layout is often smoothed, and thus does not always accurately represent the garment shape. Using control points instead of the parser naturally avoids the issues caused by non-optimal layout and achieves optimal rendering results. 

\subsubsection{Multi-Garment Virtual Try-on}
Multi-garment try-on is more challenging than single garment because the framework needs to work with a variety of garment categories and layer multiple garments appropriately. O-VITON~\cite{Neuberger_2020_CVPR} first synthesizes a semantic layout to outline the garment interactions and then broadcast the feature encoding vectors based on the layout. However, the feature encoding vectors are not able to preserve structural patterns. POVNet~\cite{Kedan_Li_2021_CVPR} proposed an iterative method of constructing the outfit, overlaying an item on top of another at each step of the generation. This framework also warps the garment and thus is able to preserve the attributes. 

We extend the multi-garment try-on framework by allowing users to style an outfit on a person in different ways. We also improve the results by being able to better coordinate multiple garments during the rendering process, for example ensuring that the garment beneath does not stick out of the garment on top (Figure~\ref{fig:outerwear_skirt}).

\begin{figure*}
\begin{center}
\vspace{-0.5cm}  
	\includegraphics[width=0.98\linewidth]{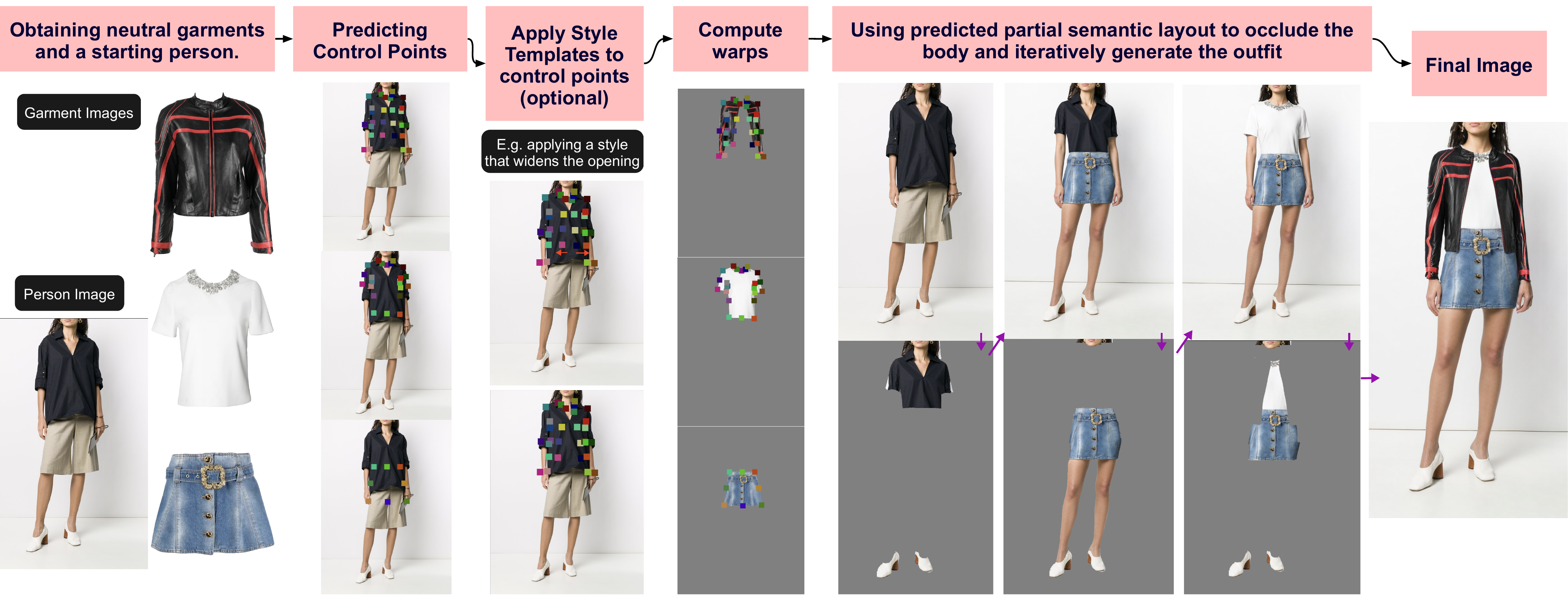}
    \caption{This figure shows an overview of our outfit generation pipeline. We first predict the control points of each garment based on a neutral garment image and the body pose of the person. Then, we can modify the drape of each garment by applying certain style templates to adjust the control points. Subsequently, we predict each garment warp based on the edited control points. Finally, we generate each garment on the person sequentially based on the layering (garments beneath are generated first). During the process, the predicted semantic layout is used to occlude the specific region of the image that we are generating for each garment.} 
    \label{fig:inference_process_overview}
    \end{center}
\vspace{-0.7cm}  
\end{figure*}

\subsection{Fashion Edits}

Zhu \etal~\cite{zhu2017be} allows editing an outfit by a simple text description through a conditional GAN. Fashion++~\cite{hsiao2019fashionplus} proposed a framework to perform minimal edit to a specific fashion item to maximize the fashionability of an outfit. Han \etal~\cite{Han2019CompatibleAD,
  inpainting-based} enable edits to the shape and appearance of the garment through an inpainting method. SwapNet~\cite{Raj2018SwapNetIB} transfers the style of the garments from one person to another. Dong
\etal~\cite{Dong_2020_CVPR} enables editing to the person and garment through occluding a region and filling in with color strokes and sketches, and synthesizing using a conditional normalization method. Cui\etal~\cite{cui2021dressing} recently proposed a recurrent generation pipeline to sequentially dress garments on a person. Chen\etal~\cite{Chen_2021_ICCV} allows the viewpoint of try-on to
be controlled through a sequence of poses.

Unlike the above works that mostly achieve style edits by modifying the properties of the garment or the body pose, we allow the user to edit how the garment is worn on an identical body while ensuring that the garment identical is preserved after the edit. Thus, we make use of control points as the editing medium to allow changes to the drapes and deformation of the garment but not the visual attributes.

\section{Controllable Outfit Visualization}

A typical VTON pipeline starts with predicting the semantic layout to determine where the garment and body are positioned; then, a warp is predicted to align the garment onto the body; finally, an image generator takes in the warped garment, the layout, and other features to synthesize the final image. To control the output of this pipeline, there needs to be a reliable way to control the semantic layout generator and the warper. 

\begin{figure*}
\begin{center}
\vspace{-0.7cm}
	\includegraphics[width=0.98\linewidth]{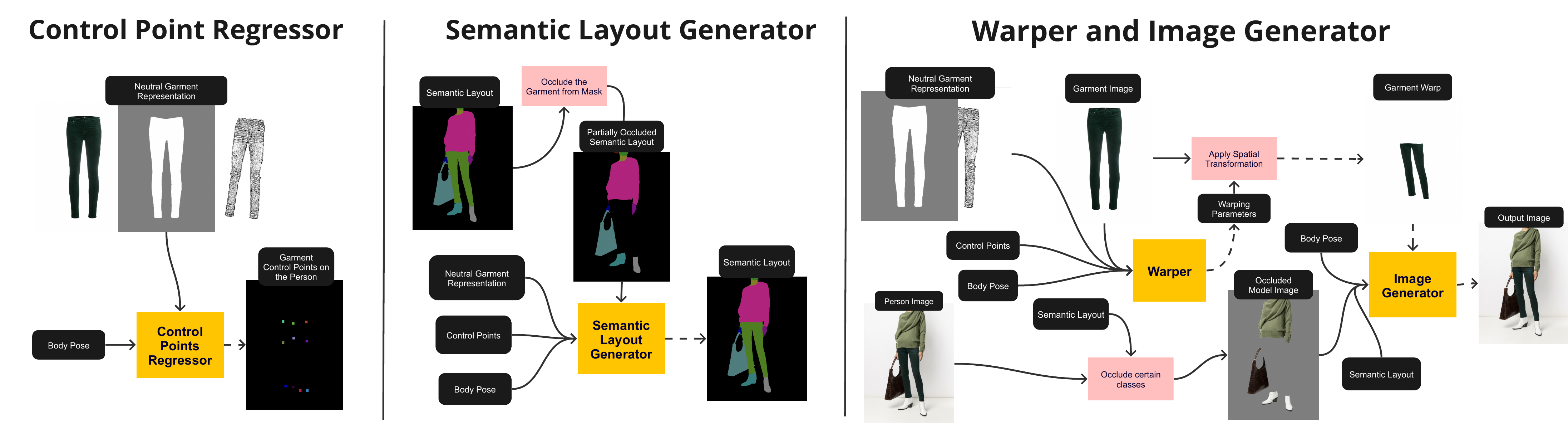}
    \caption{The figure shows an overview of the training process. The Control Point Regressor $R_c$ first learns to predict the control points based on body pose and neutral garment features. The Semantic Layout Generator $G_L$ learns to reconstruct the partially occluded layout using the control points and other features. The Warper $W$ predicts a warp to align the neutral garment onto the body based on the control points and the body pose. The predicted warp and other features are fed into the Image Generator $G_I$ to synthesize the output image. The Warper is trained jointly with the Image Generator. The dashed lines indicate the path of backpropagation during training.
   }    
    \label{fig:training_pipeline}
    \end{center}
\vspace{-0.7cm}  
\end{figure*}

Our method uses a set of garment control points $K$ on the person to enable control. We first train a Control Point Regressor $R_c$ to predict the control points $K$ from neutral garment image features $A$ and body pose keypoints $b_p$. Then, we train the semantic layout generator $G_L$ and the warper $W$ to make predictions conditioned on these control points $K$, as outlined in Figure~\ref{fig:training_pipeline}. The advantage of this setup is that we are able to move these control points around during inference to alter the garment accordingly. The way the information flows in the pipeline makes it natural for the rest of the pipeline to follow the control points. Please find the complete set of notations and definitions in the Appendix.

During inference, we first run the control points regressor to produce a different set of control points $K$ for each garment. 
A typical VTON framework can only generate one way to wear a particular set of garments.
However, our method can produce different ways of wearing the garments by using edited control points throughout the pipeline to control the drape of the garments. The garments will behave according to the edits, as shown in Figure~\ref{fig:continuous_split}.

\subsection{The Control Points}
\label{label:control_points}

The control points $K$ is a set of 2D coordinates on the person image $\{ k_1, k_2, ..., k_n \}, k_i = \left(  x_i, y_i \right)$ each having a specific semantic meaning (e.g., left shoulder, right inner sleeve, etc..). Using control points to guide the warp and the layout is desirable because it is intuitive to understand how these edits would change the garment. For example, increasing the distance of the control points between two sides of a jacket will widen the split of the jacket, as in Figure~\ref{fig:continuous_split}. It is also important that the control points embed garment semantics. Having such a property allows us to make edits on an instance of garment and apply the edits to other garments of the same category.

The DeepFashion dataset already contains garment semantic key points annotation on the human body ~\cite{DeepFashion2}. We took their pre-trained network and predicted the key points for every model image $b$ in our training dataset. The original key point annotation has redundancy (e.g., there are separate sets of collar key points for t-shirt and shirt). We merge the key points with identical semantic meaning and obtain a list of 49 unique control points $K$ (details in the Appendix).

\subsection{Training Procedure}

The system consists of the Control Points Predictor $R_c$, the Semantic Layout Generator $G_L$, the Warper $W$, and the Image Generator  $G_I$, as shown in Figure~\ref{fig:training_pipeline}.
 
\subsubsection{Control Points Regressor}

The Control Points Regressor $R_c$ takes in the body pose representation $b_p$, the neutral garment features $A$ and the control variables $Z$, and outputs the garment control points $K\prime = R_c \left(A, b_p, Z \right)$. $Z = [z_1, z_2,...,z_n]$ is a control vector where each index can control a discrete style (e.g., $z_1$ is set to $0$ for full tuck and $1$ for untuck). The number of learned styles can vary based on what annotation the dataset contains. During training, $Z$ is provided with the ground truth style values; during inference, $Z$ can be configured to control the style of the predictions.  

$R_c$ consists of a ResNet32~\cite{He2016DeepRL} connected to a fully-connected layer. $Z$ is broadcasted to a 2D plane and concatenated with other inputs before feeding them into $R_c$. The output of the fully-connected layer is reshaped into $N\times n \times2$ where $N$ is the batch size and $n$ is the number of control points. The network is trained using a $\mathcal{L}_1$ loss and $\mathcal{L}_2$ loss computed between $K$ and $K\prime$. In addition, we compute a structural consistency loss $\mathcal{L}_s$ that penalized errors in pair-wise distance between the control points. This is important because we are more concerned about the structure of the garment being misrepresented or distorted over minor displacement of the entire garment. We compute a matrix of the distance between each pair of points \[
D =
\begin{bmatrix}
    d(k_1, k_1) & d(k_1, k_2) & \dots  & d(k_1, k_n) \\
    d(k_2, k_1) & d(k_2, k_2) & \dots  & d(k_2, k_n) \\
    \vdots & \vdots & \vdots & \vdots \\
    d(k_n, k_1) & d(k_n, k_2) & \dots  & d(k_n, k_n) \\
\end{bmatrix}
\]
where $d(k_i, k_j) = \sqrt{|x_i - x_j|^2 + |y_i - y_j|^2}$, and train the network to minimize the structural consistency loss $\mathcal{L}_s = ||D - D'||$. Any rotation or translation of the garment will result in a small structural consistency loss, but anything else will yield a large loss. The total training loss for $R_c$ can be written as 
\vspace{-0.2cm} 
\begin{equation}
\mathcal{L}_{R_c} = \lambda_{1}\mathcal{L}_1 + \lambda_{2}\mathcal{L}_2 + \lambda_{3}\mathcal{L}_s
\end{equation}
where $\lambda_{1}$, $\lambda_{2}$ and $\lambda_{3}$ are the weights for each component. 

Note that the network will always predict a value for every control point but not every control point exists for a garment (e.g., tops don't need trouser leg control points). Thus, we mask out the control points that do not exist in the current garment when computing the training loss.

\subsubsection{Semantic Completion Generator}

Following most VTON pipelines~\cite{tprvton,Issenhuth2020DoNM, Kedan_Li_2021_CVPR, Chopra_2021_ICCV, Style_Based_Global_Appearance}, we train a Semantic Layout Generator $G_L$ to predict the human parsing indicating the pixel region of garments and body parts on the generated try-on image. 


To train $G_L$, we obtain the neutral garment features $A$, the model's body pose $b_p$, the garment control points $K$, and the partially occluded semantic layout mask $\hat{b_m}$. $K$ is plotted into a 2D map where each channel contains a single control point and is then concatenated with the other inputs. $G_L$'s training objective is to reconstruct the original semantic layout $b_m$ from $\hat{b_m}$. The occluded mask $\hat{b_m}$ is obtained by replacing parts of $b_m$ by background class through an occlusion function $\hat{b_m} = f_o \left(b_m, a_t \right)$. The occlusion function $f_o$ operates based on the garment category $a_t$. $f_o$ is different per garment category and guided by the following rules: (1) $f_o$ always replaces the region of the specified garment category $a_t$; (2) $f_o$ also replaces the category of skin classes that are directly connected with $a_t$. For example, when $a_t$ is a top, $f_o$ removes the arm layouts and the neckline layout, but not the legs layout. $G_L$ is trained through pixel-wise Cross-Entropy loss and adopts a U-Net architecture following prior arts~\cite{Kedan_Li_2021_CVPR}.

\subsubsection{Warper and Image Generator}

The Warper $W$ aligns $A$ to the person following the control points. The Image Generator Network $G_I$ takes in the warped garment $A^w$, the semantic layout, and other features to produce the final output image $b`$. $W$ and $G_I$ are trained jointly but applied separately during inference.

$W$ takes in the the body pose $b_p$, the neutral garment features $A$ and the control points $K$, and outputs the transformation parameters $\theta = W\left(b_p, A, K \right)$. We compute the spatial transformation through $\theta$ and obtain the warped garment features $A^w = {a^w, a^w_m, a^w_c, a^w_e}$ ($a^w$ is the warped garment image; $a^w_m$ is the warped garment mask; $a^w_c$ is the warped garment cropped mask; $a^w_e$ is the warped garment edge map.). 

 The main learning objective of $W$ is to minimize the difference in appearance between the warped garment and the region of the warp on the person. Our pipeline can use different warper implementations (e.g.  Thin-Spline Warper, Optical Flow Warper, Multiple Coordinated Affine Warper, etc..)~\cite{Han_2019_ICCV,wang2018toward,tprvton, ge2021parser, Kedan_Li_2021_CVPR, cycle_consistency_tryon, choi2021vitonhd, Chopra_2021_ICCV, Style_Based_Global_Appearance}. In our experiment, we adopt the flow warper formulation from~\cite{Chopra_2021_ICCV} but use OpenPose~\cite{8765346} instead of DensePose~\cite{Guler_2018_CVPR} as the body presentation, because Li~\etal ~\cite{Kedan_Li_2021_CVPR} demonstrated that DensePose representations are often biased by the garment on the body. We adopt the warping loss of prior work~\cite{Chopra_2021_ICCV}, written as $\mathcal{L}_w$.

The Image Generator $G_I$ produce the final try-on image $b' = G_I(A^w, b_p, b_o, b_m^g)$ based on the warped garment features $A^w$, the body pose representation $b_p$, the occluded person image $b_o$ and the input semantic mask $b_m^g = b_m \odot \left(1 - b_g\right)$ without the to try-garment's layout $b_g$. The occluded person image $b_o$ is created by applying the occluded semantic layout mask $\hat{b_m}$ produced by $f_o$ on the person image $b$. We remove the garment mask from the semantic layout $b_m$ because the garment warp may not exactly match the shape of the mask during. Removing the garment mask allows $G_I$ to figure out the garment shape through the warped garment features $A^w$ and often yield better results. This procedures was also adopted by other VTON works~\cite{Chopra_2021_ICCV, Style_Based_Global_Appearance}.

The Image Generator $G_I$ architecture is a U-Net as the skip connections provide an easy way to copy the provided input image. The learning objective is to produce an image that resembles the ground truth model and appears realistic. We train $G_I$ with an $\mathcal{L}_1$ loss and an $\mathcal{L}_{perc}$ perceptual loss~\cite{Johnson2016Perceptual} computed between $b'$ and $b$. In addition, we train $G_I$ with an adversarial loss $\mathcal{L}_{adv}$ to encourage realism of the output image following prior arts~\cite{Han_2019_ICCV,wang2018toward,tprvton, ge2021parser, Kedan_Li_2021_CVPR, cycle_consistency_tryon, choi2021vitonhd, Chopra_2021_ICCV, Style_Based_Global_Appearance, ge2021parser}. 

The combined training loss for $W$ and $G_I$ becomes
\vspace{-0.2cm} 
\begin{equation}
\mathcal{L}_{G_f} = \lambda_{1}\mathcal{L}_1 + \lambda_{2}\mathcal{L}_{perc} + \lambda_{3}\mathcal{L}_{adv} + \lambda_{4}\mathcal{L}_w
\end{equation}
where $\lambda_{1}$, $\lambda_{2}$, $\lambda_{3}$ and $\lambda_{4}$ are the weights the loss.

\subsection{Outfit Generation Procedure}

During inference, we start from a set of garments $\{ a_1, a_2, ..., a_o \}$ and a person image $b$, and generate an image of the person wearing the outfit. As outlined in Figure~\ref{fig:inference_process_overview}, we first compute the control points and the warps for every garment; then, we iteratively generate each garment on the person by applying $G_L$ and $G_I$, which highly resemble the iterative inference process of Li~\etal~\cite{Kedan_Li_2021_CVPR}. 

Predicting the control points for all the garments first is advantageous because certain types of style edits require coordinating multiple items in an outfit. Knowing the control points of all garments also allows us to coordinate them and make edits to avoid mistakes, as shown in Figure~\ref{fig:outerwear_skirt}.

\subsection{Applying the Edits to Garments}


Our method controls how garments drape by either choosing different control parameters $Z$ for the Control Point Regressor $R_c$ or directly modifying the coordinates of the predicted control points $K'$. 

$Z$ controls a set of discrete drapes/styles (e.g. tuck vs. untuck, open vs. closing the outerwear). Because it is difficult to change between two discrete styles by moving around the control points (e.g. closed outerwear requires a different set of control points), it is more convenient to have $R_c$ learn to predict them. Fortunately, discrete styles can be easily defined and labeled through heuristics. During training, we provide the discrete style labels as the control parameters $Z$, so $R_c$ learns to predict control points for the provided style. 

Directly editing $K'$ is better for styles that are continuous, such as how an opened-outerwear drapes. The simplest form of style edit can be done by applying some offsets to specific control points. For example, to wear the waistline of a skirt higher, one could shift the vertical coordinate of every control point upward by a constant, as shown in the Appendix. Other types of edits require coordinating multiple items in the outfit. For example, to achieve the front tuck in Figure~\ref{fig:different_tuck} row 4, we set the middle torso control point of the shirt to be at the same height as the  waistline control point of the bottom. 

 \paragraph{Split Outerwear} Splitting outerwear requires cutting a neutral garment into two regions and warping each region separately. In this scenario, we divide the garment representation of an outerwear into left garment $A^l$ and right garment $A^r$. The control points are also divided into left component $K^l$ and right component $K^r$. The warper predicts the spatial transformation parameter for the left side as $\theta^l= W\left(b_p, A^l, K^l \right)$ and the right side as $\theta^r= W\left(b_p, A^r, K^r \right)$. Finally, both sides of the warps are merged into a single warped image and fed into the image generator $G_I$. The way each side of the outerwear drapes is guided by the corresponding control points, as shown in Figure~\ref{fig:continuous_split}. 

 \begin{figure}
\begin{center}
\vspace{-0.2cm}
	\includegraphics[width=0.98\linewidth]{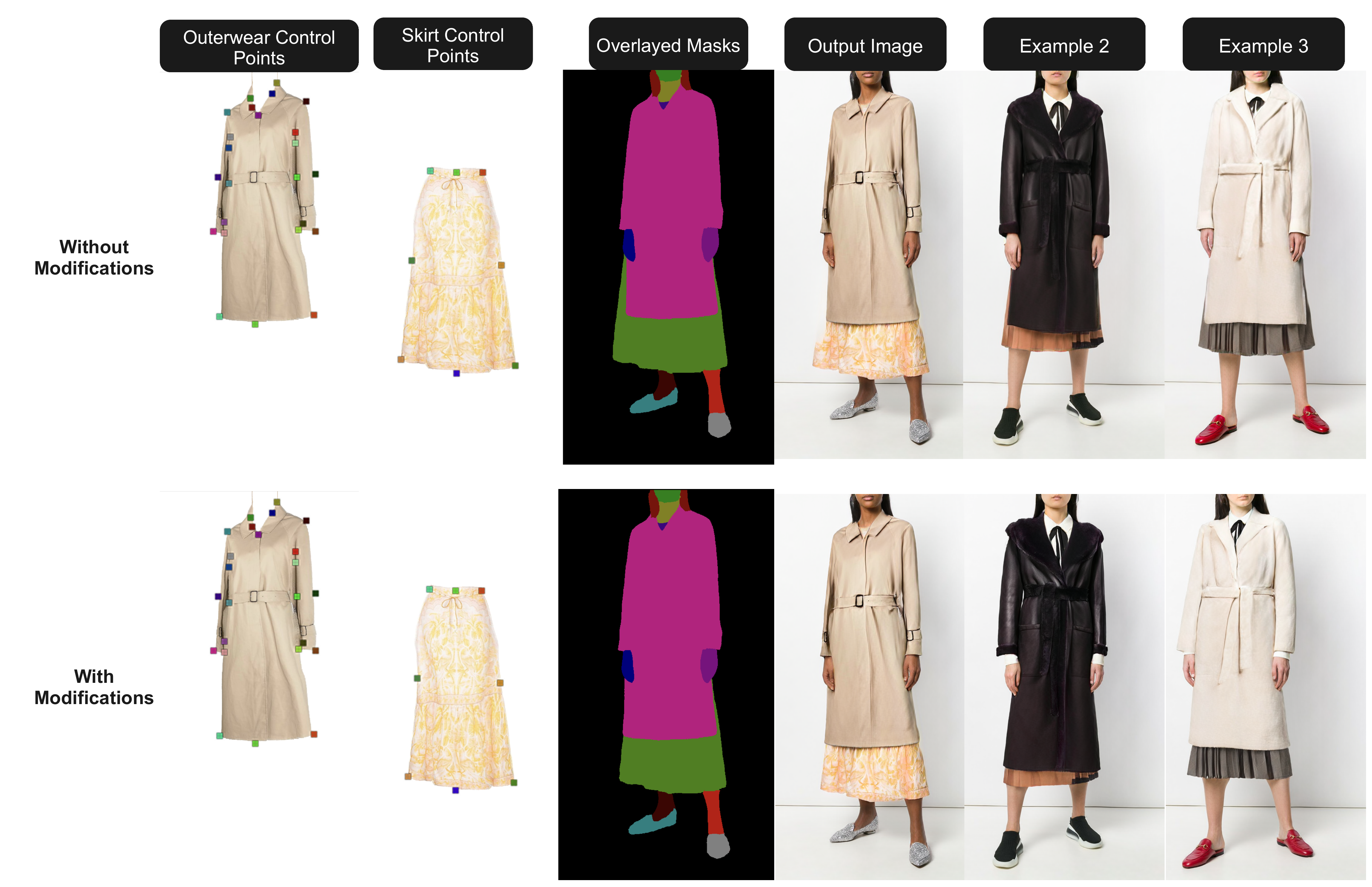}
    \caption{An example of coordinating the control points of multiple garments to avoid errors in the rendering. On the top row, the long jacket fails to cover the skirt, as they were predicted separately. We are able to address the error by adjusting the skirt's control points to be within the outerwear. This adjustment can also be applied to different outerwear skirt pairs.
  }    
    \label{fig:outerwear_skirt}
    \end{center}
\vspace{-0.5cm}  
\end{figure}
 
 \paragraph{Coordinating the Garments in an Outfit} Because the on-body control points of each individual garment are predicted separately, they may not coordinate well. Figure~\ref{fig:outerwear_skirt} shows an example of this error when the predicted region of the skirt unnaturally sticks out of the coat. These types of errors can be easily addressed by modifying garment control points. In the above example, the system addresses the error by shifting the position of the skirt to fall within the coat region. This adjustment allows the outerwear to completely cover the skirt and shows how the control points can be used to coordinate multiple garments to prevent artifacts. 

\section{Experiments}

\subsection{Dataset Preparation}

We train our network on the OVNet dataset~\cite{Kedan_Li_2021_CVPR}, which contains a variety of garments of different categories worn in different styles. To obtain control points, we first apply a pre-trained garment keypoints prediction network from DeepFashion2~\cite{DeepFashion2} on the person images in the dataset and then convert them into control points using the procedure described in Section~\ref{label:control_points}. To obtain style labels for $Z$, we use simple heuristics to label the styles: for closed vs. open outerwear, we check whether the outerwear has two large disjoint regions of similar size in the layout; for tuck vs. untuck, we threshold the height of waistline garment control points against the waist height in the body pose.

\subsection{Quantitative Evaluation}

We compare our method with state-of-the-art outfit visualization methods to demonstrate that our method yields optimal generation quality. We evaluate the performance using the Frechet Inception Distance (FID)~\cite{heusel2017gans} because it doesn't require ground truth data (the OVNet dataset only has one style per outfit, whereas we are generating different styles). We obtained a test set with 8 different styles as shown in Table~\ref{table:quant}. For each style, we sample 10k real images using a combination of heuristics and compare them against 10k generated images of the same style.

Results in Table~\ref{table:quant} show that our method outperforms the state-of-the-art methods on the combined dataset (includes all images) as well as for each style-specific subset. This demonstrates that our method yields higher quality images while enabling control. Note that the FID score for only dresses is significantly lower than other styles that have multiple garments. This suggests that the garments-interactions of a generated look still have notable differences from a real look. The FID score for all methods on the combined dataset is lower than on individual styles, suggesting that outfit generation to match a specific style is a more challenging task. 


\begin{table}[!t]
\renewcommand{\arraystretch}{1.3}
\caption{This table compares the FID Score~\cite{heusel2017gans} between our method and state-of-the-art outfit visualization methods. Method 1 (M1) is the original implementation of OVNet~\cite{Kedan_Li_2021_CVPR}. Method 2 (M2) is the OVNet framework combined with the Flow Warper from ~\cite{Chopra_2021_ICCV}. We compute the result on the complete data set and on each individual style. (t is top; b is bottom; o is outerwear) }
\label{table:quant} 
\centering

\begin{tabular}{cc|c|c|c}

\hline
Gender &  Style & M1 & M2 & Ours \\ 
\hline
Female & dress & 15.6 & 16.2 & \textbf{14.2}  \\
Female & t+b untuck & 26.4 & 21.3 & \textbf{19.5}  \\
Female & t+b tuck & 22.4 & 22.6 & \textbf{20.2} \\
Female & t+b+o closed & 21.2 & 21.7 & \textbf{19.4}  \\
Female & t+b+o open & 23.9 & 20.8 & \textbf{19.0} \\
Male & t+b & 20.3 & 20.9 & \textbf{18.8}  \\
Male & t+b+o closed & 21.5 & 19.5 & \textbf{19.1}  \\
Male & t+b+o open & 22.1 & 18.0 & \textbf{ 17.3}  \\
\hline
\multicolumn{2}{c|}{Combined}  & 13.9 & 12.6 & \textbf{11.2}  \\
\hline
\end{tabular}
\vspace{-0.2cm} 
\end{table}

\subsection{Qualitative Evaluation}

\begin{figure}
\begin{center}
\vspace{-0.2cm}
	\includegraphics[width=0.98\linewidth]{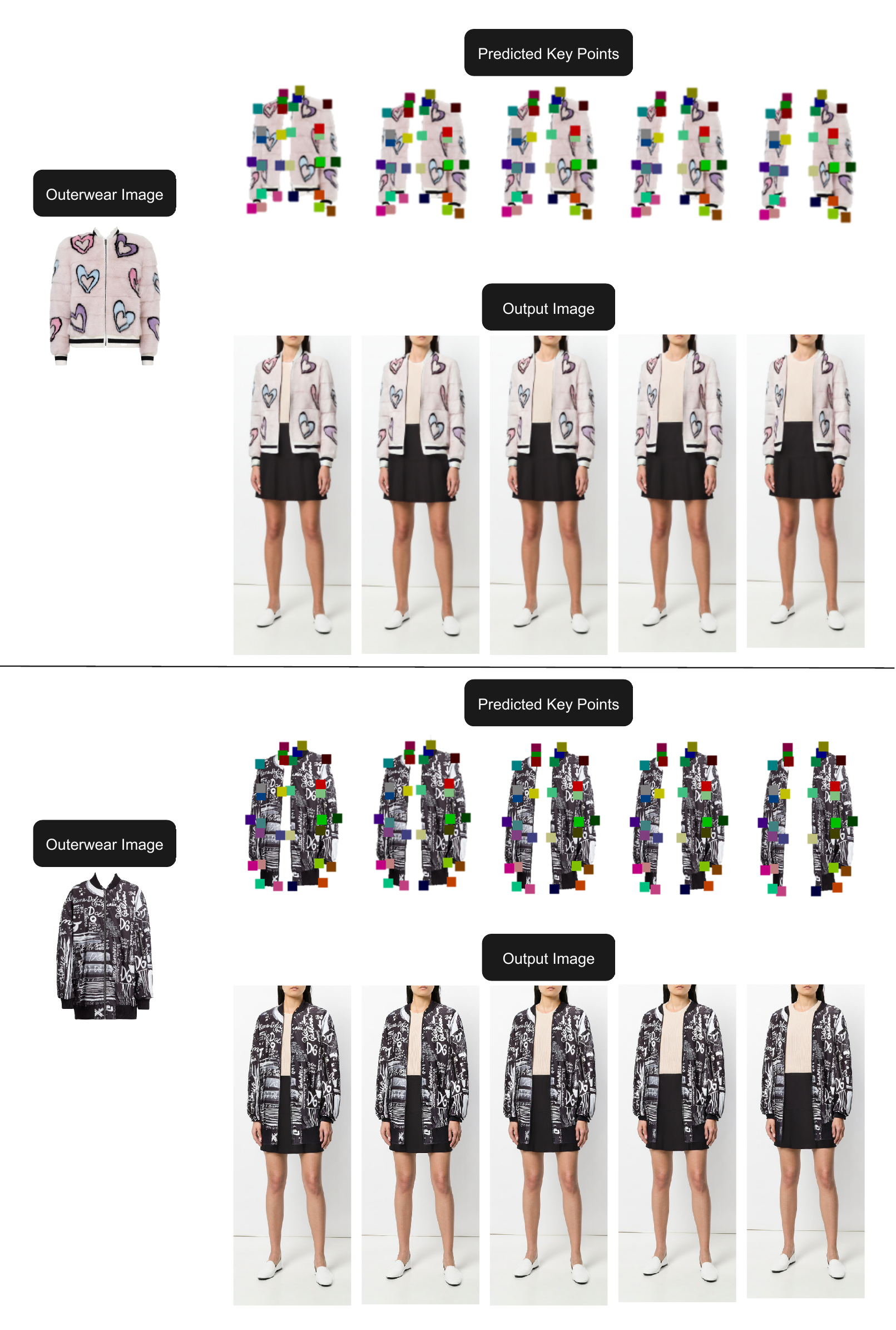}
    \caption{ This figure shows an interpolation of the control points. We move the control points by a small offset every step to gradually open the outerwear. The drape of the outerwear closely follows the control point in the entire process, demonstrating that the control points are highly effective in controlling the garment.} 
    \label{fig:continuous_split}
    \end{center}
\vspace{-0.5cm}  
\end{figure}

In Figure~\ref{fig:different_tuck} and ~\ref{fig:different_outerwear}, we demonstrate a variety of style edits that our system can support. Example edits include untucking or tucking the top in different ways, wearing the outerwear closed or open with different kinds of drapes, wearing bottoms at different waistline heights, etc. In addition, we show that each of the style edits applies consistently to many garments in the same category and appears consistent even when the garments have different properties (e.g., the length and shape of the outerwear differ). This is made possible by using garment control points that capture category-level garment semantics.

The qualitative examples also show the properties of the garments are preserved after the style edits and are not impacted by the way garments are worn. 
Pay attention to the red blocks (column 6), the colored hearts (column 7), the street painting (column 8), and the leopard prints (column 9) in Figure~\ref{fig:different_outerwear}. These distinct features are preserved when the outerwear is open or altered in different ways.

To evaluate how the control points affect the results, we perform interpolation by moving the control points slightly after each step and comparing the differences. As shown in Figure~\ref{fig:continuous_split}, the silhouette of the outerwear exactly follows the changes to the control point. The results show that the control points in our pipeline are very effective at controlling how a garment drapes.

In addition, qualitative examples show our method can control a specific garment without affecting other garments in the outfit. As shown in Figure~\ref{fig:different_outerwear}, the method only intends to edit the outerwear and preserves the look of the other garments. The visible parts of the top and the bottoms remain identical, besides the addition of some shading that is necessary to produce a realistic look. This property is desirable as one may only want to alter one item in a look.

\subsection{User Studies}
We performed user studies to evaluate our ability to preserve garment identity and consistently apply style edits (see Appendix for details).

{\bf In the first study}, we show subjects pairs of outfits rendered from our method and ask the subjects if a garment in the pair of outfits is identical. The questions contain rendered pairs of the same garment worn in different ways and rendered pairs of different garments that were chosen to look visually similar. We asked 44 respondents to answer 22 questions, where 15 were pairs of the same garment and 7 were pairs of different garments. The respondents had an overall accuracy of 85.0\%, which suggests the subjects believed that most style edits did not alter garment identity. 

{\bf In the second study}, we show the subjects a rendered outfit and ask the person to choose the style that best describes the look from a set of options. For example, a subject may be asked to choose whether a top+bottoms outfit is either untucked, fully-tucked, front-tucked, side-tucked, or half-tucked. Prior to answering the questions, subjects are shown examples of each style option. We asked 44 respondents to answer 22 styling questions. The respondents had an overall accuracy of 80.1\%, indicating that subjects are able to identify the style in most cases. Note that some of the styles are sometimes difficult to distinguish because the definitions are not binary (e.g. draping vs. split outerwear). 

\section{Conclusion \& Discussion}
We propose the first outfit try-on system that allows garments to be worn in different ways while producing try-on images that preserve state-of-the-art quality. Because our control layer embeds garment semantics, our method allows style edits made on an example garment to be applied to style other garments of the same category. Evaluation results show that the control mechanism can effectively alter the way a garment drapes while preserving its identity. The system provides a powerful tool for users to experiment with outfit drapes and for professionals to style different garments appropriately. Moving forward, we hope to extend the controllability to creating dynamic body poses and supporting different body sizes.

{\small
\bibliographystyle{ieee_fullname}
\bibliography{egbib}
}

\end{document}